\documentstyle[mncite,psfig]{mn}

\def\et{\em et al. \em}

\def\gte{\,\lower.6ex\hbox{$\buildrel >\over \sim$} \, }
\def\lte{\,\lower.6ex\hbox{$\buildrel <\over \sim$} \, }

\journal{Submitted to MNRAS}

\title{The Milky Way is just an average spiral}

\author[S.\,P.\,Goodwin \et]{Simon\,P.\,Goodwin$^1$,  John Gribbin$^1$,  Martin\,A.\,Hendry$^2$\\
$^1$ Astronomy Centre, University of Sussex, Falmer, Brighton, BN1 9QH \\
$^2$ Department of Physics and Astronomy, University of Glasgow, Glasgow G12 8QQ}

\begin{document}

\maketitle

\begin{abstract}

It has been thought for many years that the Milky Way is an overly 
large spiral galaxy.  Using Cephied distances to 17 spiral galaxies 
we calculate the true linear diameters of those galaxies.  These 
diameters are then compared to that of the Milky Way which is 
found to be, at most, an averagely sized spiral galaxy.  When compared to 
galaxies of approximately the same Hubble type ($2<T<6$) the Milky Way 
is found to be slightly undersized.  This suggests that the Hubble parameter 
is at the lower end of the currently accepted range of possibilities.

\end{abstract}


\section{Introduction}

It is part of astronomical lore that we inhabit an overly large spiral 
galaxy.  This belief has little grounding but remains as a hang-over from 
early large estimates of the value of the Hubble parameter ($H_0 > 100$ km 
s$^{-1}$ Mpc$^{-1}$) which implied that the Milky Way was perhaps the 
largest spiral galaxy in the observable Universe (Hubble 1936).  There is 
still considerable uncertainty about the value of $H_0$, and values as 
high as 80 km s$^{-1}$ Mpc$^{-1}$ and as low as 50 km s$^{-1}$ Mpc$^{-1}$ 
have recently appeared in the literature (Freedman 1994; Sandage 
1996). When the linear diameters of distant galaxies are measured 
from their angular diameters and distances obtained solely from their 
redshifts, high values of $H_0$ make galaxies closer and therefore 
smaller.  Such a situation goes against the cosmological prejudice that 
applies the principle of terrestrial mediocrity: that there is nothing 
special about where or when we live and observe from (Vilenkin 1995).  We 
seem to live on an ordinary planet orbiting an ordinary star, and it is 
natural to infer that the Solar System resides in an ordinary galaxy.

The diameter of a galaxy is problematic to define and here we  
take it to be the face-on diameter of the 25 B-mag arcsec$^{-2}$ 
isophote, allowing direct comparison between external galaxies and the 
Milky Way.  We present calculations of the true 25 
B-mag arcsec$^{-2}$ isophotal diameters of 15 spiral galaxies with 
independent distance estimates derived from Cepheid variable observations, 
mostly carried out in the past few years with the Hubble Space Telescope, 
and compare these with the inferred diameter of the Milky Way at this 
same surface brightness.

\section{The diameters of other galaxies}

Until recently, it has not been possible to compare the size of the Milky 
Way with the sizes of a statistically meaningful sample of 
other nearby spirals because very few independent distance estimates 
had been obtained to such galaxies. This situation has now changed 
dramatically with the advent of the Hubble Space Telescope. The external 
galaxies in our sample were chosen because they have distances that have 
been determined via the application of the Cepheid period-luminosity relation, 
which has long been recognised as the most reliable primary extragalactic 
distance indicator. The availability of an accurate independent distance 
estimate removes any requirement to assume a Hubble parameter or  
correct for any peculiar motion. 17 calibrating galaxies were chosen, 
mainly from the targets of the HST `Key Project' survey (Kennicutt, Freedman 
\& Mould 1995) whose distances have recently been collated in the 
literature (Giovanelli 1996; Freedman 1996).  

All of these galaxies are included in the RC3 bright galaxy catalogue from 
which their Hubble type ($T$) and isophotal diameters ($D_{25 {\rm (ang)}}$) 
were taken (de Vaucoulers et al. 1991).  These isophotal diameters have been 
corrected for Galactic extinction but not for inclination as the RC3 
catalogue assumes that the discs are optically thick and hence the major 
axis diameter is used directly.  

Table 1 presents the angular diameters, distances and inferred linear 
diameters of 17 spiral galaxies.  

\begin{center}
\begin{table}
\begin{center}
\begin{tabular}{|c|c|c|c|c|c|} \hline
NGC & M & T & $D_{25 {\rm (ang)}}$ & $d$ & $D_{25 {\rm (true)}}$ \\
 & & & arcmin & Mpc & kpc \\ \hline
224  & 31  & 3  & 204  & 0.77 & 45.7 \\
300  &     & 7  & 22.4 & 2.15 & 14.0 \\
598  & 33  & 6  & 74.1 & 0.85 & 18.4 \\
925  &     & 7  & 11.0 & 9.38 & 30.0 \\
1365 &     & 3  & 11.2 & 18.2 & 59.3 \\
2366 &     & 10 & 8.32 & 3.44 & 8.32 \\
2403 &     & 6  & 22.9 & 3.18 & 21.2 \\
3031 & 81  & 2  & 27.5 & 3.63 & 29.0 \\
3109 &     & 9  & 20.0 & 1.23 & 7.16 \\
3351 &     & 3  & 7.59 & 10.1 & 22.3 \\
3368 & 96  & 2  & 7.59 & 11.6 & 25.6 \\
3621 &     & 7  & 13.5 & 6.80 & 26.7 \\
4321 & 100 & 4  & 7.59 & 16.1 & 36.2 \\
4496 &     & 9  & 3.98 & 16.8 & 19.5 \\
4536 &     & 4  & 7.59 & 16.7 & 36.8 \\
4639 &     & 4  & 2.82 & 25.1 & 20.6 \\
5457 & 101 & 6  & 28.8 & 7.38 & 61.8 \\ \hline
\end{tabular}
\end{center}
\caption{The Hubble type $T$, face-on angular 25 B-mag arcsec$^{-2}$ 
isophotal diameters $D_{25 {\rm (ang)}}$, Cephied distances $d$ and actual 
25 B-mag arcsec$^{-2}$ isophotal diameters $D_{25 {\rm (true)}}$ for 
17 spiral galaxies.}
\end{table}
\end{center}

\section{The size of the Milky Way}

The 25 B-mag arcsec$^{-2}$ isophotal diameter of the Milky Way has been 
calculated by assuming that the Galactic disc is well represented by an 
exponential disc (Freeman 1970) and using the surface brightness 
profile equation from de Jong (1996)

\begin{equation}
r = \frac{h(\mu(r) - \mu_0)}{1.086}
\end{equation}

\noindent where $\mu_0$ is the central surface brightness of a galaxy and 
$h$ is the disc scale length.  For the Milky Way $\mu_0 = 22.1 \pm 0.3$ 
B-mag arcsec$^{-2}$ and $h = 5.0 \pm 0.5$ kpc (van der Kruit 1987, 1990).  
This leads to a derived 25 B-mag arcsec$^{-2}$ isophotal diameter for the 
Milky Way of 

\[
D_{25{\rm (true)}} = 26.8 \pm 1.1 {\rm kpc}
\]

In order to test the validity of the above formula we applied it to the 
two galaxies in the calibrating sample for which values of $h$ and 
$\mu_0$ were available to us: M31 (van der Kruit 1990) and M100 (van der 
Kruit 1987).  The $D_{25}$ diameter for M31 was found to be 43.2 kpc 
(5.5\% below  the RC3 value) and that for M100 was found to be 36.3 kpc 
(0.3\% above the RC3 value).  This gives us confidence that the assumption of 
an exponential disc is justified and reasonable.

There is some uncertainty as to the Hubble type of the Milky 
Way.  Evidence appears to favour a classification of the Milky Way as 
an Sbc galaxy ($T = 4$), however morphologies between Sab and Scd ($T = 2$ 
to 6) cannot be ruled out (van der Kruit 1987, 1990). There has also been 
recent evidence indicating the presence of a triaxial bar-like structure in 
the central region of the Milky Way, with a major axis scale length of the 
order of 1 kpc (Stanek et al. 1994). It seems unlikely that the existence 
of such a bar would significantly bias the determination of the disc scale 
length for the Milky Way quoted above, although we intend in future work to 
carry out a more detailed comparison of disc sizes in barred and 
unbarred galaxies.

\section{Conclusion}

Figure 1 shows a histogram of the distribution of spiral galaxy sizes for 
all of the 18 galaxies in our sample, with the positions of the Milky Way and 
M31 indicated.  The Milky Way lies almost exactly on the mean of the 
galaxy sizes (actually, just below the average as 
$<D_{25 {\rm true}}> = 28.3$ kpc).  

It is even more interesting to compare the Milky Way with galaxies of a 
similar Hubble type.  Figure 2 shows the histogram obtained for the 12 
galaxies of Hubble types 2 through 6.  In this case the Milky Way lies 
further below the average linear diameter of 33.6 kpc; one should not read 
too much into this, however, since the Milky Way still lies well within 
one standard deviation of the sample mean. A more quantitative statistical 
analysis would clearly require a larger calibrating sample and also a 
more realistic model for the distribution of linear diameters (Sodre \& 
Lahav 1993).

There seems no doubt, however, that the Milky Way is {\em not} one of 
the largest spiral galaxies.  NGC 1365 and NGC 5457 (M101), in particular,  
are the local giants, more than twice as large as the Milky Way.  This 
confirms Eddington's (1933) prescient comment, made more than 60 years ago 
that the `relation of the Milky Way to the other galaxies is a subject upon 
which more light will be thrown by further observational research, and 
that ultimately we shall find that there are many galaxies of a 
size equal to and surpassing our own.'

The implications of this conclusion for estimates of the Hubble Parameter 
are clear - specifically, if the Milky Way is an average spiral this may 
favour estimates of $H_0$ at the lower end of the range of accepted values.    
We are carrying out a detailed analysis of the implications for $H_0$ which 
will be published shortly.  

\vspace{0.5cm}

\noindent {\bf Aknowledgements}

The authors acknowledge the use of the Starlink computers at the Universities 
of Sussex and Glasgow.

\newpage

\begin{figure*}
\centerline{\psfig{figure=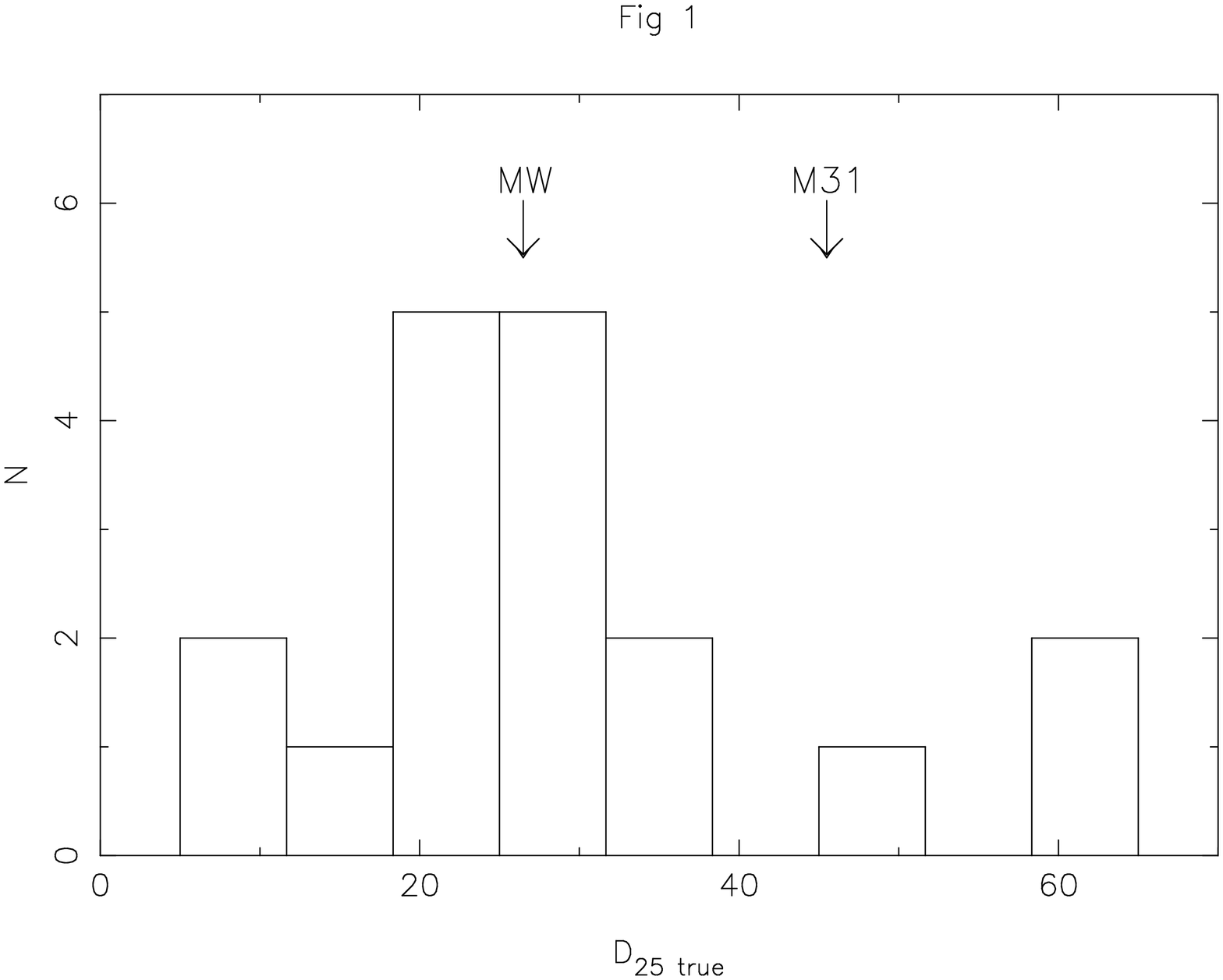,height=10.0cm,width=16.0cm,angle=270}}
\caption{A histogram of the true diameters of all 18 spiral 
galaxies in the sample.  The diameters of the Milky Way (MW) and M31 have 
been marked.}
\end{figure*}

\begin{figure*}
\centerline{\psfig{figure=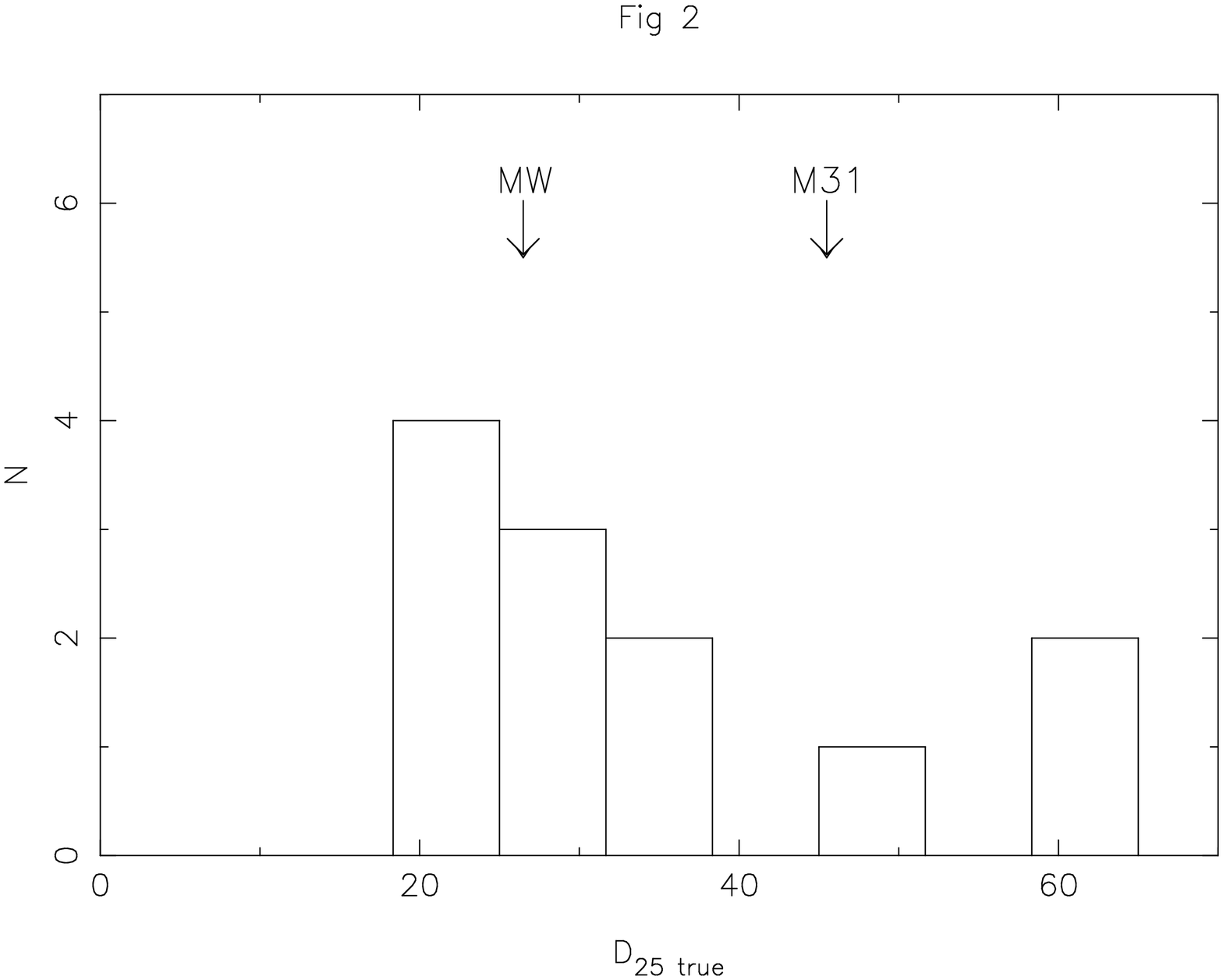,height=10.0cm,width=16.0cm,angle=270}}
\caption{A histogram of the true diameters 12 galaxies in the 
sample with Hubble types 2 to 6 (the range of possible Milky Way values), 
again the diameters of the Milky Way (MW) and M31 have been marked.}
\end{figure*}


\begin{thebibliography}{}

\bibitem[]{}Eddington\,A.\,S.  1933, `The Expanding Universe' (Cambridge 
University Press)

\bibitem[]{}Freedman\,W.\,L. et al. 1994, Nature, 371, 757 

\bibitem[]{}Freedman\,W.\,L.  1996, to appear in Critical Dialogues 
in Cosmology, ed. N.\,Turok, LANL preprint no. astro-ph/9612024

\bibitem[]{}Freeman\,K.\,C. 1970, ApJ, 160,  811

\bibitem[]{}Giovanelli\,R.  1996, to appear in The Extragalactic Distance 
Scale, eds. M.\,Livio, M.\,Donahue \& N.\,Panagia (Cambridge), LANL preprint 
no. astro-ph/9610116 

\bibitem[]{}Hubble\,E. 1936, `The Realm of the Nebulae' (Yale University Press)

\bibitem[]{}de Jong\,R.\,S.  1996, A\&A Suppl., 118, 557

\bibitem[]{}Kennicutt\,R.\,C., Freedman\,W.\,L., Mould\,J.R. 1995, AJ, 110, 
1476 

\bibitem[]{}Sandage\,A. 1996, AJ, 111, 18

\bibitem[]{}Sodre\,L. Jr., Lahav\,O. 1993, MNRAS, 260, 285

\bibitem[]{}Stanek\,K.Z. et al. 1994, ApJ, 429, L73

\bibitem[]{}van der Kruit\,P.\,C.  1987, in The Galaxy, eds. G.\,Gilmore \& 
    B.\,Carswell (NATO), p27 

\bibitem[]{}van der Kruit\,P.\,C.  1990, in The Galactic and Extragalactic 
    Background Radiation, eds. S.\,Bowyer \& C.\,Leinert (IAU), p 85

\bibitem[]{}de Vaucoulers\,G., de Vaucouleurs\,A., Corwin\,H.\,G., 
Buta\,R.\,J., Paturel\,G. \& Fouqu\'{e}\,P.  1991, Third Reference Catalogue 
of Bright Galaxies (RC3) 

\bibitem[]{}Vilenkin\,A. 1995, Phys. Rev. Lets., 74, 846

\end{thebibliography}
\end{document}